\documentclass[twocolumn,aps,amsmath,amssymb,prd,nofootinbib]{revtex4}
\usepackage{epsfig}
\usepackage{wasysym}

\def\al{\alpha}
\def\be{\beta}

\def\de{\delta}
\def\ep{\epsilon}

\def\et{\eta}

\def\ka{\kappa}
\def\la{\lambda}
\def\rh{\rho}

\def\ta{\tau}

\def\ph{\phi}

\def\om{\omega}
\def\Ga{\Gamma}

\def\Ph{\Phi}

\def\mn{{\mu\nu}}

\def\fr#1#2{{{#1} \over {#2}}}
\def\frac#1#2{{\textstyle{{#1}\over {#2}}}}

\def\lsim{\mathrel{\rlap{\lower4pt\hbox{\hskip1pt$\sim$}}
    \raise1pt\hbox{$<$}}}
\def\gsim{\mathrel{\rlap{\lower4pt\hbox{\hskip1pt$\sim$}}
    \raise1pt\hbox{$>$}}}

\def\prt{\partial}

\def\etal{{\it et al.}}

\def\pt#1{\phantom{#1}}

\def\sb{\overline{s}}
\def\tb{\overline{t}}
\def\ub{\overline{u}}
\def\kf{(k_F)}
\def\cf{(c_F)}

\def\htr{\overline{h}}
\def\xx'{|\vec x -\vec x'|}
\def\con#1#2#3{\Ga^{#1}_{\pt{#1}#2#3}}

\newcommand{\beq}{\begin{equation}}
\newcommand{\eeq}{\end{equation}}
\newcommand{\bea}{\begin{eqnarray}}
\newcommand{\eea}{\end{eqnarray}}
\newcommand{\rf}[1]{(\ref{#1})}

\begin{document}

\title{Lorentz-violating gravitoelectromagnetism} 

\author{Quentin G.\ Bailey}

\affiliation{Physics Department,
Embry-Riddle Aeronautical University,
3700 Willow Creek Road,
Prescott, AZ 86301, USA}

\date{May 2010; Physical Review D, in press}

\email{baileyq@erau.edu}

\begin{abstract}
The well-known analogy between a special limit 
of General Relativity and electromagnetism 
is explored in the context of the 
Lorentz-violating Standard-Model Extension (SME).  
An analogy is developed for the minimal SME that connects 
a limit of the CPT-even component of the 
electromagnetic sector to the gravitational sector. 
We show that components of the 
post-newtonian metric can be directly 
obtained from solutions to the electromagnetic sector. 
The method is illustrated with specific examples including
static and rotating sources.
Some unconventional effects that arise for 
Lorentz-violating electrostatics and magnetostatics 
have an analog in Lorentz-violating post-newtonian gravity.
In particular, 
we show that even for static sources, 
gravitomagnetic fields arise in the presence of Lorentz violation.
\end{abstract} 

%\pacs{11.30.Cp, 04.25.Nx, 03.30.+p}

\maketitle

\section{Introduction}
\label{introduction}

In its full generality, 
General Relativity (GR) is a highly 
nonlinear theory that bears little resemblance to
classical Maxwell electrodynamics.
Nonetheless, 
it has long been known that 
when gravitational fields are weak, 
and matter is slow moving,
analogs of the electric and 
magnetic fields arise for gravity 
\cite{thirring}.
These fields are sourced by a scalar density 
and vector current density,
just as in electrostatics and magnetostatics.
Furthermore, 
in the geodesic equation for a test body,
terms of no more than linear order in the velocity resemble the classical
Lorentz force law arising from effective gravitoelectric
and gravitomagnetic fields 
\cite{lt}.
Also, 
a well-known analogy exists between 
the precession of classical spin 
in a gravitational field
and the precession of the spin of
a charged particle in an electromagnetic field 
\cite{ps,thomas}.

In this work we investigate 
the fate of the standard connection 
between stationary solutions 
of the Einstein and Maxwell theories
when violations of local Lorentz symmetry are introduced.
Recent interest in Lorentz violation has been 
motivated by the possibility of uncovering
experimental signatures from an underlying unified theory 
at the Planck scale \cite{strings,cpt,LVreviews}.
We examine the modified Einstein and Maxwell equations provided by 
the action-based Standard-Model Extension (SME) framework, 
which allows for generic Lorentz violation
for both gravity and electromagnetism,
among other forces \cite{sme1,sme2,akgrav}.

In the so-called minimal SME case, 
the electromagnetic sector
contains $23$ observable coefficients
for Lorentz violation organized into two parts:
$4$ CPT-odd coefficients 
in $(k_{AF})^\mu$ with dimensions of mass
and $19$ CPT-even coefficients in the 
dimensionless $\kf^{\mu\nu\ka\la}$ \cite{sme2,em1}.
The former set has been stringently constrained
by astrophysical observations at the level 
of $10^{-42}\, {\rm GeV}$ \cite{kaf,nonmin}.
The latter set has been explored over the last 
decade using astrophysical observations \cite{em2}
and sensitive laboratory experiments including
resonant-cavity tests \cite{rescav}, 
among others \cite{kf}.\footnote{For a thorough list
of experiments, see the collected data tables in Ref.\
\cite{tables}.}
Currently,
constraints on these $19$ coefficients 
are at the level of $10^{-14}$ to $10^{-32}$ \cite{tables}.
 
In the minimal SME gravitational sector, 
there are $20$ coefficients for Lorentz violation
organized into a scalar $u$, 
two-tensor $s^\mn$, 
and four-tensor $t^{\mu\nu\ka\la}$ \cite{akgrav}.
Within the assumption of spontaneous Lorentz-symmetry breaking, 
the dominant effects for weak-field gravity
are controlled by the subset called $\sb^\mn$ \cite{qbkgrav}.
These gravity coefficients have been explored so far
in lunar laser ranging \cite{llr} and
atom interferometry \cite{atom,atom2,atom3}, 
while possibilities exist for other tests \cite{bsl},
including time-delay and Doppler tests \cite{qgrav}.

Since the CPT-even portion of 
the electromagnetic sector of the minimal SME
has $19$ coefficients and the gravitational sector, 
apart from an unobservable scaling $\ub$, 
also has $19$ coefficients,
one might expect a correspondence between the two sectors
- an extension of the conventional analogy.
Indeed, 
as we show in this work, 
there {\it is} a correspondence under certain restrictions. 
Thus it turns out that, 
under certain circumstances, 
Lorentz violation affects classical electromagnetic systems
in flat spacetime in a similar manner as gravitational systems
are affected by Lorentz violation in the weak-field limit of gravity.
As a consequence, 
some of the unusual effects that occur for 
Lorentz-violating electromagnetism
have an analog in the gravitational case.
In addition,
from a practical perspective, 
it is quite useful to be able to translate analytical
results in one sector directly into the other,
as we illustrate toward the end of this work.

We begin in Sec.\ \ref{field equations} 
by reviewing the basic field equations for
the gravitational and electromagnetic sectors of the SME. 
Next we explore the solutions to these equations
and establish the analogy between the two sectors
in both the conventional case and 
the Lorentz-violating case in Sec.\ \ref{gems field match}.
In Sec.\ \ref{test-body motion},
we explore test body motion for both sectors
and establish the connection in 
the conventional and Lorentz-violating cases.
We conclude this work in Sec.\ \ref{examples and applications}
by illustrating the results with 
the examples of a pointlike source and 
a rotating spherical source, 
and we discuss some experimental applications
of the results.
Finally in Sec.\ \ref{summary}, 
we summarize the main results of the paper. 
Throughout this work, 
we take the spacetime metric signature to be $-+++$
and we work in natural units where $c=\ep_0=\mu_0=1$.

\section{Field equations}
\label{field equations}

The CPT-even coefficients for Lorentz violation
in the photon sector of the minimal SME
are denoted $\kf^{\mu\nu\ka\la}$, 
which is assumed totally traceless by convention, 
and have all of the tensor symmetries of the Riemann
tensor and therefore contain 
$19$ independent quantities \cite{sme2,em1}.
Following Ref.\ \cite{nonmin}, 
it is useful to split these $19$ 
coefficients into two independent pieces 
using the expansion
\bea
\kf^{\mu\nu\ka\la} &=& C^{\mu\nu\ka\la}
+ \fr 12 [\et^{\mu\ka} \cf^{\nu\la}
-\et^{\mu\la} \cf^{\nu\ka}
\nonumber\\
&&
-\et^{\nu\ka} \cf^{\mu\la}
+\et^{\nu\la} \cf^{\mu\ka}].
\label{kfsplit}
\eea
With this decomposition
$9$ coefficients are contained in the traceless 
combinations 
$\cf^\mn=\kf^{\mu\al\nu}_{\pt{\mu\al\nu}\al}$
and $10$ coefficients 
are in $C^{\mu\nu\ka\la}$, 
which is traceless on any two indices.
The modified Maxwell equations can then be written in 
the form 
\bea
\prt_\mu F^{\mu\nu} + C^{\mu\nu\ka\la} \prt_\mu F_{\ka\la}
\nonumber\\
+ \cf^{\nu\la}\prt_\mu F^\mu_{\pt{\mu}\la}
+ \cf^{\mu\la}\prt_\mu F_\la^{\pt{\la}\nu} 
&=& -j^\nu,
\label{maxwell}
\eea
where $F_\mn=\prt_\mu A_\nu - \prt_\nu A_\mu$
and $A_\mu$ is the vector potential.
This result follows directly from the 
electromagnetic action of the minimal SME in Minkowski spacetime, 
when the electromagnetic field is coupled in 
the standard way to a conserved four-current 
$j^\mu=(\rh, \vec J)$, 
and when the coefficients are treated as constants
in an observer inertial frame.

In the gravitational sector, 
the coefficients for Lorentz violation 
are expressed in terms of three independent
sets of coefficients: $t^{\mu\nu\ka\la}$, $s^\mn$, $u$.
The $t$ coefficients are taken as totally traceless
and have the symmetries of the Riemann curvature 
tensor, implying $10$ independent quantities.
The $s$ coefficients are traceless and 
contain $9$ independent quantities.
With the scalar $u$, 
there are in general $20$ independent
coefficients describing 
Lorentz violation in the gravitational sector.

Unlike the SME in Minkowski spacetime, 
it is not straightforward to proceed directly
from the gravitational action to the field equations.
This is because introducing externally prescribed 
coefficients for Lorentz violation into the action 
can generally conflict with the fundamental
Bianchi identities of pseudo-riemannian geometry
\cite{akgrav}.
It turns out, 
however, 
that spontaneous breaking of Lorentz symmetry
evades this difficulty \cite{akgrav,spont}.
In Ref.\ \cite{qbkgrav}, 
the linearized gravitational field equations
were derived using a formalism that treats the coefficients
for Lorentz violation as dynamical fields
inducing spontaneous breaking of Lorentz symmetry, 
with certain restrictions placed on their dynamics.\footnote{
This formalism 
can be considered a subset of the 
gravity sector of the SME expansion, 
and was recently dubbed the ``Bailey-Kosteleck\'y formalism'' 
\cite{vector}.} 
Similar methods can be adopted for the matter-gravity couplings
as well \cite{tkgrav}.
The linearized equations in this formalism include, 
as special cases, 
models of spontaneous Lorentz-symmetry breaking with scalar
\cite{scalar}, 
vector \cite{vector}, 
and two-tensor fields \cite{cardinal,phon}.

In linearized gravity the metric is expanded as
\beq
g_\mn = \et_\mn + h_\mn.
\label{linear}
\eeq
Within the minimal SME approach, 
the linearized field equations can be written in
terms of the vacuum expectation values of 
the coefficients for Lorentz violation, 
denoted $\tb^{\mu\nu\ka\la}$, 
$\sb^\mn$, 
$\ub$, 
which are taken as constants in a special
observer coordinate system.\footnote{The reader is warned not 
to confuse the bar notation which indicates the vacuum expectation values 
of the tensor fields $t^{\mu\nu\ka\la}$, 
$s^\mn$, 
and $u$ with the bar notation $\htr_\mn$ used in this paper 
for the trace-reversed metric fluctuations.}
The linearized field equations take the form
\bea
G_\mn &=& 8\pi G_N (T_M)_\mn 
+\sb^{\ka\la} R_{\ka\mu\nu\la} 
- \sb^\ka_{\pt{\ka}\mu} R_{\ka\nu}
\nonumber\\
&&
-\sb^\ka_{\pt{\ka}\nu} R_{\ka\mu} 
+\fr 12 \sb_\mn R 
+\et_\mn \sb^{\ka\la} R_{\ka\la},
\label{einstein}
\eea
where $G_N$ is Newton's gravitational constant.
In this expression $R_{\ka\mu\nu\la}$
is the Riemann curvature tensor, 
$G_\mn$ is the Einstein tensor, 
$R_\mn$ is the Ricci tensor, 
and $R$ is the Ricci scalar.
All curvature tensors in \rf{einstein} 
are understood as linearized in the fluctuations $h_\mn$.
Since the $\ub$ coefficient only scales the left-hand side, 
it is unobservable and is discarded for this work.

Because of a tensor identity \cite{qbkgrav}, 
the $10$ coefficients $\tb^{\mu\nu\ka\la}$
vanish from the linearized equations, 
thus leaving the $9$ coefficients in $\sb^\mn$ in this limit.
This immediately implies that, 
should an analogy exist between the photon
and gravity sectors of the SME, 
it involves a subset of the
$(k_F)^{\mu\nu\ka\la}$ coefficients.
This subset is comprised of the $9$ coefficients 
$\cf^\mn$.

\section{Field match}
\label{gems field match}

\subsection{Conventional GR case}
\label{conventional gr case}

In GR and Maxwell electrodynamics, 
the analogy between certain components of
the metric fluctuations $h_\mn$ and $A_\mu$ reveals itself 
from the field equations in the harmonic gauge:
\beq
\prt^\mu \htr_\mn = 0.
\label{lg}
\eeq
Here $\htr_\mn$ are the usual trace-reversed metric fluctuations
defined by 
\beq
\htr_\mn = h_\mn - \fr 12 \et_\mn h^\al_{\pt{\al}\al}.
\label{htr}
\eeq
In the absence of the coefficients for Lorentz violation
$(k_F)^{\mu\nu\ka\la}$ and $\sb^\mn$, 
the Einstein equations in this gauge read
\beq
\Box \htr_\mn = -16\pi G_N (T_M)_\mn,
\label{gr}
\eeq
while the Maxwell equations, 
in the gauge $\prt^\mu A_\mu=0$, 
are
\beq
\Box A_\mu = -j_\mu.
\label{maxwell2}
\eeq

To match the structure of the Maxwell 
equations one typically 
makes a slow motion assumption for the matter source.
For example, 
for perfect fluid matter with ordinary velocity
$v^j$ much less than one,
and small pressure, 
\bea
(T_M)_{00} &\approx& \rh,
\nonumber\\
(T_M)_{0j} &\approx& -\rh v^j, 
\nonumber\\
(T_M)_{jk} &\approx& \rh v^j v^k. 
\label{pf}
\eea
Thus, 
examining Eq.\ \rf{gr}, 
it can be seen that the components
$\htr_{jk}$ will be one power of 
velocity more than $\htr_{0j}$, 
and hence negligible. 
To be more precise, 
if one adopts the standard 
post-newtonian expansion and counts
terms in powers of mean velocity ${\overline v}$,
labeled as $O(1)$, 
$O(2)$, 
etc.,
one finds from Eq.\ \rf{gr} that 
\bea 
\htr_{00} &\sim& O(2),
\nonumber\\
\htr_{0j} &\sim& O(3),
\nonumber\\
\htr_{jk} &\sim& O(4).
\label{pnexp}
\eea

Furthermore, 
in post-newtonian counting, 
partial time derivatives obey the post-newtonian
counting \cite{sw,cmw}
\beq
\fr {\prt}{\prt t} \sim \fr {\overline v}{\overline r},
\label{partial}
\eeq
where $\overline r$ is the mean distance.
A consistent approximation including
up to $O(3)$ terms would 
take $\Box \approx {\vec \nabla}^2$ and Eq.\ \rf{gr}
would become 
\beq
{\vec \nabla}^2 
\htr_{0\mu} = -16\pi G_N (T_M)_{0\mu},
\label{gr2}
\eeq
which can be compared with the stationary equations
for electrostatics and magnetostatics
\beq
{\vec \nabla}^2 A_\mu = -j_\mu.
\label{ems}
\eeq
From these two expressions it is clear that,
given solutions for $A_\mu$ in the stationary limit, 
the solutions $\htr_{0\mu}$ can be obtained
in the manner below.
\begin{enumerate}
\label{grmaxwell}

\item Replace charge density $\rh_q$ with mass density $\rh_m$ and 
electric current density $J^j$ with mass-current density $\rh v^j$.

\item Write down the metric components as 
$\htr_{0\mu}=-16\pi G_N A_\mu$.

\end{enumerate}
This method agrees with standard results in 
the literature \cite{cw,reviews}.

\subsection{Lorentz-violating case}
\label{lorentz-violating case}

Equations \rf{gr} and \rf{maxwell2} lead to a direct correspondence
between the solutions for $\htr_{0\mu}$ and $A_\mu$.
In the presence of Lorentz violation, 
this direct analogy involving 
the trace-reversed metric fluctuations
disappears because the coefficients $\sb^\mn$ in the 
modified equations \rf{einstein} generally 
mix the components of $\htr_{0\mu}$ with $\htr_{jk}$.
As a result of this mixing, 
$\htr_{jk}$ contains terms of $O(2)$ in post-newtonian counting,
in contrast to the GR case \rf{pnexp}, 
and so there is no particular utility in using the 
trace-reversed metric fluctuations $\htr_\mn$ over
the metric fluctuations $h_\mn$. 

We focus on the stationary limit, 
where a match between the electromagnetic and gravity sectors 
can be obtained for the metric components 
$h_{00}$ and $h_{0j}$.
This gravitoelectromagnetic correspondence is
most easily obtained directly from the stationary 
solutions to Eqs.\ \rf{maxwell} and \rf{einstein} 
for the metric $g_\mn$ and the vector potential $A_\mu$.
The gravitational solutions were obtained in Ref.\ \cite{qbkgrav}
while the results in electrodynamics were obtained in 
Refs.\ \cite{qbkems,ems2}.

Before displaying the solutions here, 
it will be convenient to introduce
various potential functions that 
take a similar form for both 
the electromagnetic and gravitational sectors.
The key source quantities appearing in these potentials are the 
charge (mass) density $\rh$
and the charge (mass) current $J^j$.
The needed potentials are
\bea
U &=& \al \int \fr{\rh (\vec x')}{\xx'} d^3x',
\nonumber\\
U^{jk} &=& \al \int \fr{\rh (\vec x') (x-x')^j (x-x')^k}
{\xx'^3} d^3x',
\nonumber\\
V^j &=& \al \int \fr{J^j (\vec x')}{\xx'} d^3x',
\nonumber\\
X^{jkl} &=& \al \int \fr{J^j (\vec x') (x-x')^k (x-x')^l}
{\xx'^3} d^3x'.
\label{pots}
\eea
In the stationary limit, 
all partial time derivatives of the potentials
vanish.
The density $\rh$ is time independent and 
the current is transverse, 
$\prt_j J^j=0$.  
This implies some simplifications of the
identities among the potentials listed in 
Ref.\ \cite{qbkgrav}, 
including $\prt_j V^j=0$ and $\prt_j X^{jkl}=0$.

The electromagnetic potentials are obtained
by interpreting $\rh$ as charge density, 
$J^j$ as a steady-state current density, 
and letting the constant $\al=1/4\pi$.
For the gravitational sector, 
the potentials are obtained by interpreting
$\rh$ as mass density, 
$J^j=\rh v^j$ as mass-current density and 
letting $\al=G_N$. 

The components of the metric fluctuations $h_{0\mu}$, 
relevant for comparison with the electromagnetic sector, 
can be obtained after an appropriate coordinate gauge choice.
We choose coordinates such that
\bea
\prt_j h_{0j} &=& 0, 
\nonumber\\
\prt_k h_{kj} &=& \fr 12 \prt_j (h_{kk}-h_{00}), 
\label{gauge}
\eea
and the metric fluctuations are time independent.
To post-newtonian $O(3)$, 
the metric components $h_{0\mu}$ are then given by
\bea
h_{00} &=& (2+3\sb^{00})U + \sb^{jk} U^{jk} -4 \sb^{0j} V^j,
\nonumber\\
h_{0j} &=& -\sb^{0j}U -\sb^{0k}U^{jk} 
- 4 (1+\frac 12 \sb^{00})V^j
\nonumber\\
&&
+2 \sb^{jk} V^k + 2 \sb^{kl} (X^{klj}-X^{jkl}),
\label{gemsmetric}
\eea
where $\al=G_N$ is chosen in the expressions \rf{pots}.
Although they are not relevant for the match between 
the two sectors, 
for completeness, 
the remaining components of the metric $h_{jk}$ are given by
\bea
h_{jk} &=& [(2-\sb^{00})U +\sb^{lm} U^{lm}]\de^{jk} 
-\sb^{jl}U^{lk}
\nonumber\\
&&
-\sb^{kl}U^{lj}
+ 2 \sb^{00} U^{jk},
\label{hjk}
\eea
which is valid to post-newtonian $O(2)$.

In the electromagnetic sector, 
we choose the stationary limit and
adopt the $U(1)$ gauge condition $\prt_j A^j=0$.
The modified Maxwell equations have the solutions
\bea
A^0 &=& [1+\frac 12 \cf^{00}]U_E + \frac 12 \cf^{jk} U_E^{jk} - \cf^{0j} V_E^j
\nonumber\\
&&
-C^{0j0k} U_E^{jk}-C^{0jkl} X_E^{ljk},
\nonumber\\
A^j &=& \frac 12 \cf^{0j}U_E +\frac 12 \cf^{0k}U_E^{jk}+ 
[1-\frac 12 \cf^{00}]V_E^j
\nonumber\\
&&
-\frac 12 \cf^{jk} V_E^k - \frac 12 \cf^{kl} [X_E^{klj}-X_E^{jkl}],
\nonumber\\
&&
-C^{0kjl} U_E^{kl}
-C^{0j0k} V_E^k
-C^{jklm} X_E^{mkl},\nonumber\\
\label{maxwellsolns}
\eea
where the subscript $E$ reminds us to take $\al=1/4\pi$ in 
the potentials \rf{pots}.

A glance at Eqs.\ \rf{gemsmetric} and \rf{maxwellsolns}
reveals that many of the same terms occur in both sectors.
However, 
in the electromagnetic sector the contributions
from the $10$ independent coefficients $C^{\mu\nu\ka\la}$
do not vanish.
To match the two sectors we must first restrict our attention to 
the special case where
\beq
C^{\mu\nu\ka\la}=0.
\label{cond1}
\eeq
Next we split the terms appearing in $A_\mu$
and $h_{0\mu}$ into those involving potentials 
derived from charge density $\rh$ 
and those derived from current density $J^j$.
These fields are defined as 
\bea
(h_{\rh})_{00} &=& (2+3\sb^{00})U + \sb^{jk} U^{jk},
\nonumber\\
(h_J)_{00} &=& -4 \sb^{0j} V^j,
\nonumber\\
(h_{\rh})_{0j} &=& -\sb^{0j}U -\sb^{0k}U^{jk}, 
\nonumber\\
(h_J)_{0j} &=& - 4 (1+\frac 12 \sb^{00})V^j
+2 \sb^{jk} V^k 
\nonumber\\
&&
+ 2 \sb^{kl} (X^{klj}-X^{jkl}),
\nonumber\\
(A_{\rh})^0 &=& [1+\frac 12 \cf^{00}]U_E + \frac 12 \cf^{jk} U_E^{jk}, 
\nonumber\\
(A_J)^0 &=& - \cf^{0j} V_E^j,
\nonumber\\
(A_{\rh})^j &=& \frac 12 \cf^{0j}U_E +\frac 12 \cf^{0k}U_E^{jk},
\nonumber\\
(A_J)^j &=& [1-\frac 12 \cf^{00}]V_E^j
-\frac 12 \cf^{jk} V_E^k 
\nonumber\\
&&
- \frac 12 \cf^{kl} [X_E^{klj}-X_E^{jkl}].
\label{split}
\eea
Note that the split of $A_\mu$ and $h_{0\mu}$
corresponds to splitting the terms in the post-newtonian
metric into $O(2)$ and $O(3)$ and splitting the terms in the 
electromagnetic potentials into ``post-coulombian'' terms of
$O(2)$ and $O(3)$ \cite{cmw,ll}. 
The correspondence between the two sectors is summarized in 
Table \ref{gemsmatch}.

\begin{table}
\begin{center}
\begin{tabular}{|l|c|c|}
\hline
Quantity & Electromagnetic sector & Gravitational sector \\
\hline
Coefficients & $\cf^\mn$ & $\sb^\mn$ \\
Scaling & $1/4\pi$ & $G_N(1+\sb^{00})$ \\
Density $\rh$ & charge density & mass density \\
Current $J^j$ & current density $J^j$ & mass current $J^j=\rh v^j$ \\ 
$\rh$ fields  & $(A_{\rh})_\mu$ & $(h_{\rh})_{0\mu}$ \\
$J^j$ fields & $(A_J)_\mu$ & $(h_J)_{0\mu}$ \\
\hline
\end{tabular}
\caption{\label{gemsmatch}
The gravitoelectromagnetic correspondence between
the electromagnetic and gravitational sectors
of the minimal SME.}
\end{center}
\end{table}

Given a stationary solution to the modified Maxwell
equations \rf{maxwell} in the Coulomb gauge ($\prt_j A^j=0$), 
one can obtain the corresponding metric components
by using the following procedure.
\begin{enumerate}
\label{gemsmap}
\item Set $C^{\mu\nu\ka\la}=0$.

\item Replace $\cf^\mn \rightarrow \sb^\mn$.

\item Separate $A^\mu$ into density-sourced and current-sourced 
terms $(A_\rh)_\mu$ and $(A_J)_\mu$.

\item Replace charge density $\rh_q$ with mass density $\rh_m$ and 
electric current density $J^j$ with mass-current density $\rh v^j$.

\item Write down the metric components 
\bea
(h_{\rh})_{0\mu} &=& -8\pi G_N (1+\sb^{00})(A_\rh)_\mu,
\nonumber\\ 
(h_J)_{0\mu} &=& -16\pi G_N (1+\sb^{00}) (A_J)_\mu,
\label{smegems}
\eea
and omit any subleading order terms ($O(\sb^2)$).

\end{enumerate}

The close resemblance of the effects of Lorentz violation on gravity 
and electromagnetism is remarkable considering the qualitative 
differences between the theories, 
particularly in the starting lagrangians and field equations \cite{akgrav}.
On the other hand, 
since there is a known analogy between 
$A_\mu$ and $h_{0\mu}$ in the conventional case, 
and both sectors are affected by two-tensor coefficients 
for Lorentz violation, 
one might have expected a close correspondence
in the appropriate limit.
In fact, 
the map constructed above further justifies the 
construction of the post-newtonian metric using the formalism
in Ref.\ \cite{qbkgrav}, 
which itself relied on several assumptions concerning the 
dynamics of spontaneous Lorentz-symmetry breaking.

An interesting feature of the solutions for Lorentz-violating 
electrodynamics is the mixing of electrostatic and magnetostatic 
effects in the stationary limit.  
As can be seen from \rf{split}, 
this occurs because a part
of the scalar potential $A^0$ depends on current density 
and part of the vector potential $\vec A$ depends on 
charge density, 
a feature absent in the conventional case.
This was aptly named electromagnetostatics (EMS) in Ref.\ \cite{qbkems}.
For Lorentz-violating gravity, 
a similar mixing occurs and $h_{00}$ depends partly on mass current
while $h_{0j}$ depends partly on mass density, 
resulting in what can be called gravitoelectromagnetostatics (GEMS).
These features are illustrated with specific examples 
in Sec.\ \ref{examples and applications}.

Note that other possibilities are open for exploration 
concerning the match between the two sectors of the SME. 
For example, 
we do not treat here the interesting possibility
of whether an analogy persists
using gravitational and electromagnetic tidal tensors, 
as occurs in the Einstein and Maxwell
theories \cite{ch}.

\section{Test-body motion}
\label{test-body motion}

In this section we study another aspect 
of gravitoelectromagnetism.
This concerns the behavior of matter in
the presence of the stationary 
gravitational or electric and magnetic fields.
As we show below,
if one adopts the appropriate limit, 
the behavior of test masses in gravitational
fields and test charges in electric and magnetic
fields is analogous, 
despite the presence of Lorentz violation.
However, 
differences do arise in the presence of
Lorentz violation when comparing the 
gravitational spin precession to the classical spin precession 
of a magnetic moment in the presence of
electromagnetic fields.

\subsection{Geodesic motion}
\label{geodesic motion}

When Lorentz violation is present in the 
electromagnetic sector only, 
test charges $e$ obey
\beq
\fr {du^\mu}{d\ta} = \fr em F^\mu_{\pt{\mu}\nu} u^\nu,
\label{forcelaw}
\eeq
where $u^\mu$ is the four-velocity.
With the usual identification of the electric and
magnetic fields, 
$E_j=F_{j0}$ and $B_j=(1/2)\ep_{jkl}F_{kl}$, 
we can write the spatial components of \rf{forcelaw}
as the familiar Lorentz-force law:
\beq
\fr {du^j}{dt} = \fr em [E^j+(\vec v \times \vec B)^j]
\label{forcelaw2}
\eeq
For small velocities, 
$u^j\approx v^j=dx^j/dt$.
Thus, 
with $k_F$ affecting only the electromagnetic sector, 
the force law for charges is conventional \cite{em1}.

In the SME, 
restricted to only the $\sb^\mn$ coefficients,
freely falling test bodies satisfy the usual
geodesic equation
\beq
\fr {du^\mu}{d\ta} = - \Ga^\mu_{\pt{\mu}\al\be}u^\al u^\be.
\label{geodesic}
\eeq
In its full generality, 
the structure of \rf{geodesic} is quite different
from Eq.\ \rf{forcelaw} for charges.
Nonetheless, 
in the weak-field slow motion limit of gravity, 
there is a correspondence.

Changing variables in \rf{geodesic} to coordinate time, 
one can solve for the coordinate acceleration
$a^j = dv^j/dt$ in terms of the connection
coefficients projected into space and time components
using standard methods.
One obtains the well-known expression \cite{sw},
\bea
a^j &=& -\con j00 - 2 \con j0l v^l -\con jkl v^k v^l
\nonumber\\
&&
+\left(\con 000 +2 \con 00k v^k +\con 0kl v^k v^l\right) v^j.
\label{geodesic2}
\eea

So far, 
Eq.\ \rf{geodesic2} is an exact result, 
and bears little resemblance to Eq.\ \rf{forcelaw2}.
If one then assumes that the test particle velocity is 
small and keeps only terms linear in the test particle velocity $v^j$, 
the acceleration becomes
\beq
a^j = -\con j00 - 2 \con j0k v^k + \con 000 v^j.
\label{geodesic3}
\eeq
To get a match with equation \rf{forcelaw2}
additional assumptions are needed.
For example, 
in the post-newtonian approximation, 
the dominant contributions to the 
connection coefficients are given by the formulas
\bea
\con j00 &=& \prt_0 g_{0k} - \fr 12 g^{jk} \prt_k g_{00},
\nonumber\\
\con j0k &=& \fr 12 \prt_0 g_{jk} + 
\fr 12 (\prt_k g_{0j}-\prt_j g_{0k}),
\nonumber\\
\con 000 &=& -\fr 12 \prt_0 g_{00},
\label{connections}
\eea
which is valid to post-newtonian $O(4)$.
If the metric is stationary in the chosen coordinate system, 
($\prt_0 g_\mn=0$), 
the acceleration, 
in terms of the metric fluctuations $h_\mn=g_\mn-\et_\mn$,
is given by
\bea
a^j &=& \fr 12 \prt_j h_{00} 
+v^k (\prt_j h_{0k}-\prt_k h_{0j})
\nonumber\\
&&
- \fr 12 h_{jk} \prt_k h_{00},
\label{geodesic4}
\eea
which neglects terms proportional to the 
test-mass velocity squared but otherwise is
valid to post-newtonian $O(4)$.
This expression now resembles the Lorentz-force law \rf{forcelaw2}
except for the last nonlinear term.

To be consistent with the post-newtonian approximation
to $O(4)$, 
the last term must be included, 
as well as nonlinear contributions to 
$h_{00}$ at $O(4)$.
This is because the second term
in Eq.\ \rf{geodesic4}, 
the so-called gravitomagnetic acceleration term, 
is an $O(4)$ term in the post-newtonian expansion.

\subsubsection{GR case}
\label{gr case}

Results from GR are contained in \rf{geodesic4}
and \rf{forcelaw2} in the limit 
of vanishing coefficients for Lorentz violation.
In the stationary limit of GR, 
and in the coordinate gauge \rf{gauge}, 
the acceleration \rf{geodesic4} can be written as
\beq
a^j = \prt_j \ph 
-4 v^k (\prt_j V^k-\prt_k V^j).
\label{geodesic5}
\eeq
Here $\ph$ is a post-newtonian potential that includes
$O(4)$ terms in GR \cite{cmw}: 
\beq
\ph = \int \fr{(\rh +\rh \Pi+3p-2\rh U)}{\xx'} d^3x' -2U^2, 
\label{phi}
\eeq
where $p$ is the perfect fluid pressure and 
$\Pi$ is the internal energy per unit mass.
Note that $\ph$ does {\it not} satisfy the
field equation \rf{gr2},
\beq
{\vec \nabla}^2 
\htr_{00} = -16\pi G_N \rh.
\label{gr3}
\eeq
Instead it satisfies
\beq
{\vec \nabla}^2 \ph = -4\pi G_N 
\left(\rh +\rh \Pi+3p-2\rh U\right)-4 (\vec \nabla U)^2.
\label{phieqn}
\eeq
Therefore $\ph \neq \htr_{00}$, 
and it {\it cannot} be obtained directly from 
the solutions to $A_0$ in Eq.\ \rf{ems} using
the standard match.

Generally, 
care is required in discarding the nonlinear terms in $\ph$, 
while keeping the second, gravitomagnetic 
terms in Eq.\ \rf{geodesic5}.
A simple estimate for a realistic 
scenario can establish this.
For a rotating spherical body, 
the solution for $V^j$ is of
order $G_N I\om/r^2 \sim G_N MR^2 \om/r^2$,
where $I$ is the inertia of the body, 
$R$ its radius, 
$\om$ its angular velocity, 
and $r$ is the coordinate distance from the 
origin to the location of the test body.
The typical test particle velocity $v^j$ 
is of order $v \sim \sqrt{G_N M/r}$ or less, 
where $M$ is the mass of the source body.
Thus, 
the contribution to \rf{geodesic5}
from the gravitomagnetic force term on 
a test particle outside of the source body,
has an approximate size
\beq
|\vec a_{gm}| \sim \fr {(G_N M) R^2 \om v}{r^3}. 
\label{est1}
\eeq
The contribution from the nonlinear terms in $\ph$
to the test particle acceleration have an approximate size
$U\vec \nabla U$ or
\beq
|\vec a_{nl}| \sim \fr {G_N M}{r} \fr {G_N M}{r^2}. 
\label{est2}
\eeq
Assuming that the nonlinear contributions are much smaller
than the gravitomagnetic contributions, 
$|\vec a_{nl}|<<|\vec a_{gm}|$,
amounts to assuming
\beq
R \om v >> \fr{G_N M}{R}.
\label{est3}
\eeq
For example, 
consider a test-body near the Earth's surface. 
For this case one finds that condition \rf{est3} implies
the unrealistic condition that the test particle velocity 
must be greater than $1/2000$ of the speed of light.

In addition to the above argument, 
it is important to recall that terms of second
and higher order in the test body velocity $v^j$
were discarded in \rf{geodesic3}. 
In terms of post-newtonian counting, 
these terms make contributions to the acceleration
$a^j$ at the same order ($O(4)$) as the nonlinear terms.
One example is the term $\con jkl v^k v^l$, 
which can be shown to have an approximate size similar
to \rf{est2} in the typical post-newtonian scenario.\footnote{In fact, 
in the case of laboratory gravitational sources and test bodies, 
these velocity-squared terms may be substantially larger 
than the nonlinear terms in Eq.\ \rf{geodesic5}, 
as discussed in Ref.\ \cite{bct}.} 
Furthermore, 
it is interesting to note that an argument along the lines of the 
one presented here appeared in the original paper 
by Lense and Thirring in 1918 \cite{lt}. 
There it was emphasized that nonlinear terms must be included in 
the equations of motion, 
in addition to the gravitomagnetic force terms, 
to properly account, for example, 
for the precession of the orbital elements of the planets.
As an alternative to this reasoning, 
one can incorporate the nonlinear terms, 
such as those occuring in Eq.\ \rf{geodesic5},
to form ``Maxwell-like'' equations, 
as pursued in Ref.\ \cite{knt}.

For simplicity here we separate out the gravitomagnetic 
and gravitoelectric acceleration terms from the nonlinear terms.
Thus we write
\beq
\vec a = \vec a_{GEM} + \vec a_{NL},
\label{asplit}
\eeq
where the separate terms are given by 
\bea
\vec a_{GEM} &=& \vec E_G + \vec v \times \vec B_G,
\nonumber\\
\vec a_{NL} &\approx & \vec \nabla (\ph - U).
\label{asplit2}
\eea
Here we have identified the gravitoelectric
and gravitomagnetic fields for GR:
\bea
\vec E_G &=& \vec \nabla U,
\nonumber\\
\vec B_G &=& -4 \vec \nabla \times \vec V.
\label{ebg}
\eea

\subsubsection{Lorentz-violating case}
\label{geod lorentz-violating case}

To see if there is any resemblance for 
the Lorentz-violating case between 
the gravitational force law and the electromagnetic force law, 
we can proceed from Eq.\ \rf{geodesic3}.
Adopting the stationary limit \rf{geodesic4}, 
we restrict attention to the gravitoelectromagnetic portion 
of the acceleration which we denote $(a^\prime)^j_{GEM}$.
This acceleration is given by
\beq
(a^\prime)^j_{GEM} = \fr 12 \prt_j h_{00} 
+v^k (\prt_j h_{0k}-\prt_k h_{0j}),
\label{geodesic6}
\eeq
For a consistent expansion to first order
in the coefficients $\sb^\mn$, 
we take $h_{00}$ to $O(3)$ 
and $h_{0j}$ to $O(2)$.
This produces an acceleration to first order in the 
coefficients $\sb^\mn$ that is at most $O(3)$.

In the presence of the 
coefficients for Lorentz violation $\sb^\mn$, 
the components of the metric
from Sec.\ \ref{lorentz-violating case} are needed 
to this order:
\bea
h_{00} &=& (h_\rh)_{00}+(h_J)_{00},
\nonumber\\
h_{0j} &=& (h_\rh)_{0j}.
\label{geodgems}
\eea
Note that the expansion of $h_{0j}$ is truncated at $O(2)$ 
since this term is multiplied by a velocity ($O(1)$) and 
therefore produces an $O(3)$ term in the acceleration.

With the considerations above, 
the gravitoelectromagnetic acceleration 
can be written to $O(3)$ as
\beq
({\vec a}^\prime)_{GEM} = \vec E_G + \vec v \times \vec B_G,
\label{accel} 
\eeq
which now resembles the result in Eq.\ \rf{forcelaw2}.
The effective electric and magnetic fields are given by
\bea
E_G^j &=& \fr 12 \prt_j [(h_\rh)_{00}+(h_J)_{00}],
\nonumber\\
B_G^j &=& \ep^{jkl} \prt_k (h_\rh)_{0l}.  
\label{egbg}
\eea
This result demonstrates that in the limit
that the gravitoelectromagnetic acceleration terms are considered, 
the force on a test body takes the same form
in the electromagnetic and gravitational sectors of the SME.

To use this result in a manner consistent with 
the post-newtonian expansion, 
additional terms at $O(4)$ but at zeroth order 
in the coefficients $\sb^\mn$ need to be included in the acceleration.
Specifically, 
the total acceleration at $O(4)$ takes the form 
\beq
a^j = (a^\prime)^j + a^j_{\rm NL} 
+ v^k [\prt_j (h_J)_{0k}-\prt_k (h_J)_{0j}],
\label{totaccel}
\eeq
where $a_{\rm NL}$ is given by Eq.\ \rf{asplit2} and
the components $(h_J)_{0j}$ are taken to zeroth order in the coefficients
$\sb^\mn$.
In the limit $\sb^\mn=0$, 
this expression reduces to the standard GR result in \rf{asplit}.

\subsection{Spin precession}
\label{spin precession}

The classical relativistic behavior of a particle with a
magnetic moment $\vec \mu$ under the influence of external 
electric and magnetic fields is well known.
Consider a particle, 
such as an electron, 
with spin $\vec s$ defined by
\beq
\vec \mu = \fr {g e}{2m} \vec s.
\label{spin}
\eeq
Here, 
$e$ is the charge of the particle, 
$m$ is the mass, 
and $g$ is the gyromagnetic ratio for the particle.
We can describe the behavior of the spin
relativistically using the spin (spacelike) 
four-vector $S^\mu$ which, 
in an instantaneous comoving rest frame, 
takes the form $(S^0=0,S^j=s^j)$.
The motion of the particle is described with 
the four velocity $u^\mu$, 
which satisfies Eq.\ \rf{forcelaw}.
In addition, 
we have the identity $S^\mu u_\mu=0$.

If we ignore field gradient forces and 
nonelectromagnetic forces, 
the behavior of the classical spin four-vector $S^\mu$
is determined by the dynamical equations \cite{jackson,thomas}
\beq
\fr {dS^\mu}{d\ta} = \fr {e}{m}
\left[\fr g2 F^{\mu\nu} S_\nu + \left(\fr g2 - 1\right)
u^\mu (S_\nu F^{\nu\la}u_\la)\right].
\label{spinprec1} 
\eeq
A formula for the precession of the spin as measured
in a locally comoving reference frame can 
be obtained by projecting $S^\mu$ along 
comoving spatial basis vectors $e^\mu_{\pt{\mu}\hat j}$, 
and making use of Eqs.\ \rf{spinprec1} and \rf{forcelaw}.
With the choice of $g \approx 2$, 
the lowest order contributions 
to this precession can be written
\beq
\fr {d S_{\hat j}}{d\ta} = 
\fr {e}{m} \left[\vec S \times \vec B - \fr 12 \vec S 
\times (\vec v \times \vec E) \right]^k \de_{k \hat j}.
\label{spinprec2}
\eeq
This result holds up to order $v^2$ in the particle's 
ordinary velocity.
Furthermore, 
Eqs.\ \rf{spinprec1} and \rf{spinprec2} will still hold
in the presence of Lorentz violation in the photon sector since 
the force law takes the conventional form \rf{forcelaw2}.

The behavior of the classical spin four-vector 
in the presence of gravitational fields is given by 
the Fermi-Walker transport equation \cite{mtw}
\beq
\fr {dS^\mu}{d\ta} =  - \con \mu\nu\la u^\nu S^\la
+u^\mu (a^\nu S_\nu),
\label{spinprec3}
\eeq
where $a^\mu$ is the acceleration of the spinning body.
For comparison with the electromagnetic case, 
we assume that the spin is in free fall ($a^\mu=0$), 
and again find the spin precession along the comoving
spatial basis $e^\mu_{\pt{\mu}\hat j}$, 
a standard technique \cite{mtw,cmw}.
The resulting precession was obtained in the post-newtonian
limit for an arbitrary metric in Ref.\ \cite{qbkgrav}
and is given by
\bea
\fr {d S_{\hat j}}{d\ta} &=& 
\de_{k \hat j} S^k 
\big[ \frac 14 (v^k \prt_j h_{00}-v^j \prt_k h_{00})
+\frac 12 (\prt_j h_{0k}-\prt_k h_{0j})
\nonumber\\
&&
\pt{\de_{k \hat j} S^k }
+\frac 12 v^l (\prt_j h_{kl}-\prt_k h_{jl})\big].     
\label{spinprec4}
\eea
which is valid to post-newtonian order $O(3)$.
Since this result was derived
for an arbitrary post-newtonian metric, 
it holds for the metric in Eqs.\ \rf{gemsmetric} and \rf{hjk}
as well.
Note that the expression \rf{spinprec4} does not immediately 
match \rf{spinprec2} due to the last terms in \rf{spinprec4} 
dependent on $h_{kl}$ at $O(2)$.
However, 
a judicious choice of coordinate gauge may 
alleviate the problem, 
as we show below.

In GR, 
we can make use of the results of Sec.\ 
\ref{conventional gr case} in the harmonic gauge.
When expressed in terms of $\htr_{\mn}$ the 
GR spin precession to $O(3)$ is 
\bea
\fr {d S_{\hat j}}{d\ta} &=& 
\left[ \fr 12 \vec S \times (\vec \nabla \times \vec g)^k 
- \fr 38 \vec S \times (\vec v \times \vec \nabla \htr_{00})^k 
\right]\de_{k \hat j},\nonumber\\
\label{spinprec5}
\eea
where $g^j=\htr_{0j}$ and we have omitted contributions from 
$\htr_{jk} \sim O(4)$.
The expression \rf{spinprec5} now resembles 
the electromagnetic counterpart, 
at least up to numerical factors.
In fact, 
one can again define effective electric and magnetic fields
for gravity: $\vec E_G=(1/4)\vec \nabla \htr_{00}$, 
$\vec B_G=\vec \nabla \times \vec g$.

We next introduce Lorentz violation in the gravitational 
sector in the form of the 
post-newtonian metric \rf{gemsmetric} and \rf{hjk}.
Unlike in GR there are off-diagonal terms 
in $h_{jk}$ that cannot be eliminated by a choice
of coordinate gauge.
As a result, 
we find that the third term in \rf{spinprec4} {\it cannot} 
be reduced to a term of the form 
$\vec S \times \vec \nabla \Ph$, 
where $\Ph$ is a scalar.
Therefore it is not possible to match the form 
of the spin precession in the gravitational sector
to the electromagnetic sector of the SME, 
the latter of which takes the form \rf{spinprec2}.
Evidently, 
this is due to the important role
of the metric components $h_{jk}$ in the
general spin precession expression \rf{spinprec4}.

\section{Examples and applications}
\label{examples and applications}

In this section we illustrate the methods of 
matching electromagnetic solutions for the fields to 
gravitational solutions for the metric components.
We also demonstrate the match between the two sectors
for test-body motion.
In our examples we study both a static
pointlike source and a rotating sphere.
Finally, 
we comment on the observability of the GEMS
mixing effects in specific gravitational tests.

\subsection{Static point source}
\label{static point source}

We consider first a point charge $q$ at rest 
at the origin in the chosen coordinate system.
The potentials in the Coulomb gauge were obtained 
in Ref.\ \cite{qbkems} and are given by
\bea
A^0 &=& \fr {q}{4\pi r} 
\left[ 1+\kf^{0j0j}-\kf^{0j0k}\hat x^j \hat x^k \right],
\nonumber\\
A^j &=& \fr {q}{4\pi r} 
\left[ \kf^{0kjk}-\kf^{jk0l}\hat x^k \hat x^l \right],
\label{ptcharge}
\eea
where $\hat x=\vec x /r$ and $r=|\vec x|$.

Using the method outlined in Sec.\ \ref{lorentz-violating case}, 
we can obtain the corresponding metric components 
$h_{0\mu}$ in the fixed coordinate gauge \rf{gauge}.
First we expand the coefficients $\kf^{\ka\la\mu\nu}$
into $C$ and $c_F$ terms using \rf{kfsplit}.
Next, 
we set all of the coefficients $C=0$, 
according to step $1$.
Then we make the replacement in 
the remaining coefficients $c_F \rightarrow \sb$. 
At this intermediate stage the potentials are
given by
\bea
A^0 &=& 
\fr {q}{4\pi r} 
\left[ 1+\frac 12 \sb^{00}+\frac 12 \sb^{jk}\hat x^j \hat x^k \right],
\nonumber\\
A^j &=& \fr {q}{8\pi r} 
\left[ \sb^{0j}+\sb^{0k}\hat x^k \hat x^j \right].
\label{ptcharge2}
\eea
Since there is no dependence of the potentials
on any current density, 
for step $3$ we simply note that in Eq.\ \rf{ptcharge2}
$A^0=(A_\rh)^0$ and $A^j=(A_\rh)^j$.
We make the replacement $q \rightarrow m$ and 
multiply the potentials 
by a factor of $-8\pi G_N (1+\sb^{00})$ 
and cancel subleading order terms ($O(\sb^2)$).
This yields
\bea
h_{00} &=& \fr {2G_N m}{r} 
\left[ 1+\frac 32 \sb^{00}+\frac 12 \sb^{jk}\hat x^j \hat x^k \right],
\nonumber\\
h_{0j} &=& -\fr {G_N m}{r} 
\left[ \sb^{0j}+\sb^{0k}\hat x^k \hat x^j \right].
\label{ptmass}
\eea

In a similar manner, 
we can also obtain effective 
gravitoelectric and gravitomagnetic fields
using \rf{egbg}:
\bea
E_G^j &=& -\fr {G_N m}{r^2} \left[\hat x^j (1+\frac 32 \sb^{00}
+ \frac 32 \sb^{kl} \hat x^k \hat x^l) -\sb^{jk} \hat x^k \right],
\nonumber\\
B_G^j &=& -\fr {2G_N m}{r^2} \ep^{jkl} \sb^{0k} \hat x^l.  
\label{ptmass2}
\eea
Using these expressions the acceleration of a test mass can be 
written in the Lorentz-force law form \rf{accel}.

An interesting feature arises from this simple solution. 
In Lorentz-violating electromagnetism, 
even a static source will generate a magnetic field.
For gravity, 
the analog of this effect occurs.
For example, 
consider the scenario in which 
the coefficients $\sb^{jk} =0$.
Apart from a scaling, 
the gravitoelectric force appears conventional.
However, 
even when the source body is static,
a test body with some initial velocity $\vec v_0$ 
will experience a gravitomagnetic force.

\begin{figure}[h]
\begin{center}
\epsfig{figure=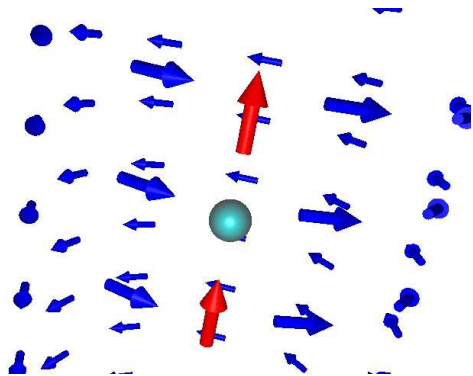,width=0.75\hsize}
\caption{\label{ptmassfig}
The gravitomagnetic field $\vec B_G$ (blue arrows)
from a static point mass $m$ (center).  
The field curls around the direction 
of $\vec s$ (red arrows) and falls off 
as the inverse square of the distance.
An approaching test body is deflected opposite $\vec s$.}
\end{center}
\end{figure}

The nature of this gravitomagnetic force
is illustrated in Fig.\ \ref{ptmassfig}.
The gravitomagnetic field itself falls off as the 
inverse square of the distance from the point mass, 
and curls around the direction of the vector 
denoted $\vec s$, 
where $s^j=-\sb^{0j}$.
A test mass approaching the pointlike source 
will be deflected in the opposite direction of $\vec s$, 
as illustrated in the figure.

\subsection{Rotating sphere}
\label{rotating sphere}

We next turn our attention to a more involved example, 
a spherical distribution of charge or mass that is rotating.
In Ref.\ \cite{qbkems}, 
a scenario was considered that involved a magnetized sphere
with radius $a$ and uniform magnetization $\vec M$.
In conventional magnetostatics, 
an idealized scenario would allow for the sphere to have 
zero charge density and no electrostatic field surrounding it, 
thus it would only produce a dipole magnetic field.
In the presence of Lorentz violation, 
however, 
a dipole electric field persists, 
with an effective dipole moment controlled by the
parity-odd coefficients for Lorentz violation $\kf^{0jkl}$.

Since we aim to find the gravitational analog of this solution, 
we cannot consider an object with zero charge density. 
Instead we study a closely related example:
a charged rotating sphere, 
which produces an effective magnetic dipole moment $\vec m$
in the conventional case.
For this example, 
the current-induced portion of the electric scalar potential, 
$(A_J)^0$,
can be obtained directly from Eq.\ (31) in Ref.\ \cite{qbkems}:
\beq
(A_J)^0 = \fr {\ep^{jkl} \cf^{0j} \hat x^k m^l}{4\pi r^2}, 
\label{emsresult}
\eeq
which holds for the region outside the sphere.
For a rotating charged sphere
\beq
m^j = \frac 13 I_E \om^j.
\label{moment}
\eeq
where $\vec \om$ is the angular velocity of the sphere.  
The quantity $I_E$ is the charge analog of the spherical 
moment of inertia for massive body,
\beq
I_E = \int d^3x \rh |\vec x|^2.
\label{ie}
\eeq

Comparing \rf{emsresult} with the standard dipole potential, 
the effective dipole moment is 
\beq
p^j= \ep^{jkl} m^k \cf^{0l}.
\label{dipole}
\eeq
The effective electric field therefore takes the standard form
\bea
\vec E = \fr {3\vec p \cdot \hat x \hat x - \vec p}{4\pi r^3}. 
\label{efield}
\eea

The gravitational analog for the solutions \rf{emsresult}
and \rf{efield} can be obtained using the methods in 
Sec.\ \ref{lorentz-violating case}.
Since the $C^{\ka\la\mu\nu}$ coefficients do not appear, 
step $1$ is redundant. 
We next make the replacement 
$\cf^{0j} \rightarrow \sb^{0j}$.
All that remains is to change $\rh_q \rightarrow \rh_m$
and multiply \rf{emsresult} by $16 \pi G_N$ which yields
\beq
(h_J)_{00} = \fr {4G_N I \ep^{jkl} \hat x^j \om^k \sb^{0l}}{3r^2}, 
\label{hj}
\eeq
where now $I$ is the spherical moment of inertia of the 
massive body, 
given by Eq.\ \rf{ie} using mass density.
Note that this produces an extra component 
of the gravitoelectric field 
$\vec E_G = (1/2) \vec \nabla (h_J)_{00}$.

In the electromagnetic case, 
part of the electrostatic field
arises from the effective current of the rotating
charged sphere, 
a feature absent in the standard Maxwell theory.
This unconventional mixing of electrostatics and
magnetostatics has an analogy for stationary 
gravitational fields produced by a rotating mass, 
in the presence of Lorentz violation. 
Thus a uniformly rotating sphere of mass
produces a gravitoelectric field whose 
strength depends on the rotation rate, 
a feature absent in standard GR. 

\begin{figure}[h]
\begin{center}
\epsfig{figure=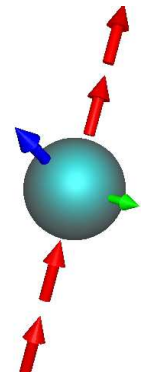,width=0.75\hsize}
\caption{\label{rotsphere}
A depiction of the effective dipole moment that develops 
for a rotating sphere in the presence of the coefficients for
Lorentz violation $\vec s$.
The dipole moment (green arrow) is proportional to the cross product 
of $\vec s$ (red arrows) with the angular momentum of the sphere
(blue arrow).}
\end{center}
\end{figure}

As in the point-mass example, 
the vector $\vec s$ is responsible for the effect.
In Fig.\ \ref{rotsphere},
the effective dipole moment of a rotating spherical mass
is depicted.
The dipole moment is obtained from the cross product of
$\vec s$ with $4 I \vec \om/ 3$.

The full solution for the case of a rotating massive or charged 
sphere can be constructed using the potentials 
$U$, 
$U^{jk}$,
$V^j$,
and $X^{jkl}$ in Eqs.\ \rf{pots}.
For the electromagnetic case ($\al=1/4\pi$) we obtain, 
for the region outside the sphere $r>R$,
\bea
U_E &=& \fr {Q}{4\pi r^2},
\nonumber\\
U_E^{jk} &=& \fr {Q \hat x^j \hat x^k}{4\pi r^2}
+ \fr {I_E}{12 \pi r^3} \left(\de^{jk}-3 \hat x^j \hat x^k \right),
\nonumber\\
V_E^j &=& \fr {I_E \ep^{jkl} \om^k \hat x^l}{12\pi r^2},  
\nonumber\\
X_E^{jkl} &=& 3V_E^j \Big[ \hat x^k \hat x^l 
\left(1-\fr {I'_E}{I_E r^2} \right) 
+ \fr {I'_E \de^{kl}}{5 I_E r^2}
\Big]
\nonumber\\
&& + \fr{I_E (\ep^{jkm}\hat x^l\om^m+\ep^{jlm}\hat x^k \om^m)}{12\pi r^2}
\left(1-\fr {3I'_E}{5I_E r^2} \right),\nonumber\\ 
\label{pots2}
\eea
where $I'_E$ is a spherical moment given by the integral 
in Eq.\ \rf{ie} with $|\vec x|^4$ instead of $|\vec x|^2$.
Using these expressions it is straightforward to calculate
the associated electric and magnetic fields
as well as the gravitoelectric and magnetic fields.
The expressions are lengthy and omitted here.

\subsection{Applications}
\label{applications}

A full analysis of the dominant observable effects in 
gravitational experiments and observations has been 
performed in Ref.\ \cite{qbkgrav}.
However, 
the coefficients were analyzed collectively 
and the separation of various distinct Lorentz-violating 
effects was not fully studied.
Here we focus specifically on the observability 
of the novel gravitomagnetic force shown to arise in 
the point-mass example in Sec.\ \ref{static point source}
and illuminate its role in a key test.

Lunar laser ranging and atom interferometry have 
measured $8$ of the $9$ coefficients in $\sb^\mn$
and the combined results are tabulated in Ref.\ \cite{atom2}.
These results are reported in the standard 
Sun-centered celestial-equatorial frame (SCF), 
where coordinates are denoted with capital letters
for clarity.
In this frame, 
the current constraints on $\sb^{JK}$
are at the $10^{-9}$ level.
For $\sb^{TJ}$, 
the constraints are at the weaker level of $10^{-6}$-$10^{-7}$.
The gravitomagnetic force due to the effective 
gravitomagnetic field in the second of Eqs.\ \rf{ptmass2} 
is controlled by the $\sb^{TJ}$ coefficients.
This force has been measured by both lunar laser ranging
and, effectively, 
atom interferometry.
However,
its specific effects are most easily discernable in 
orbital tests such as the lunar laser ranging scenario, 
so we focus on this case.

The principle effects from the $\sb^{TJ}$ coefficients 
for lunar laser ranging are modifications
to the relative acceleration of the Earth and Moon.
This acceleration includes such terms as the gravitomagnetic terms 
considered in Eqs.\ \rf{ptmass2}.
In fact, 
from the results in Ref.\ \cite{qbkgrav}, 
one can read off the portion of the Earth-Moon acceleration
$\de a^J$ responsible for the effective force that is 
described in Fig.\ \ref{ptmassfig}. 
In the SCF coordinates, 
it reads
\beq
\de a^J = \fr {2G_N \de m}{r^3} v^K (\sb^{TK}r^J-\sb^{TJ}r^K), 
\label{llr}
\eeq
where $\de m$ is the mass difference between the Earth and Moon, 
$r^J$ is the coordinate difference between the Earth and Moon 
center of mass positions and $v^J$ is their relative coordinate velocity.

The dominant observable effects from Eq.\ \rf{llr} are 
oscillations in the lunar range at the mean lunar orbital 
freqeuncy $\om$.
In the lunar laser ranging scenario, 
these oscillations are controlled by two linear combinations
of the $\sb^{TJ}$ coefficients called $\sb^{01}$ and $\sb^{02}$, 
which are expressed in the mean orbital plane of the lunar orbit.
These two quantities control the size of the 
Lorentz-violating gravitomagnetic force for this case.
Using over three decades of lunar laser ranging data, 
analysis reveals that $\sb^{01}=(-0.8\pm 1.1) \times 10^{-6}$
and $\sb^{02}=(-5.2\pm 4.8) \times 10^{-7}$ \cite{llr}.
Therefore there is no compelling evidence 
for the gravitomagnetic force controlled by $\sb^{TJ}$ coefficients.
However, 
ongoing tests such as the 
Apache Point Observatory Lunar Laser-Ranging Operation 
have already improved on lunar ranging capability
and could significantly improve sensitivity to this effect
\cite{apollo}.

\section{Summary}
\label{summary}

In this work we have shown that an analogy exists between the 
gravitational sector and the electromagnetic sector of the SME
at two levels. 
First we showed that in the stationary limit and for 
a particular coordinate choice, 
part of the post-newtonian metric $h_{0\mu}$ in 
the gravity sector can be obtained from the vector potential $A_\mu$
in the electromagnetic sector by essentially making a series 
of substitutions, 
most notably the exchange
of the coefficients $C^\mn \rightarrow \sb^\mn$, 
as outlined in Sec.\ \ref{lorentz-violating case}.
For the equations of motion of a test body, 
the gravitational case was shown to resemble the electromagnetic 
Lorentz-force law, 
so long as nonlinear terms in the geodesic equation are disregarded.

In Sec.\ \ref{examples and applications}, 
we provided two examples of how the mixing of electrostatics
and magnetostatics in Lorentz-violating electrodynamics has 
an analog in the gravitational case.
In the same manner as a point charge produces a magnetic field
in the presence of the electromagnetic coefficients $C^{0j}$, 
we showed that a point mass will produce a gravitomagnetic
field controlled by the coefficients $\sb^{0j}$.
Similarly, 
we also explored the converse of this example, 
demonstrating that a moving mass produces an additional
gravitoelectric field.
We also discussed the observability of 
the gravitomagnetic force controlled 
by the $\sb^{0j}$ coefficients in lunar laser ranging tests.

Several areas are open for future investigation.
One possibility is to systematically isolate the GEMS 
mixing effects from others in the various predicted signals 
for Lorentz violation in gravitational experiments \cite{qbkgrav}, 
along the lines of the discussion in Sec.\ \ref{applications}.
It also would be interesting to investigate whether any analogy 
is possible in the presence of the matter sector 
coefficients that play a role in gravitational experiments \cite{tkgrav,kinematics}.
Furthermore, 
using a method similar to the one developed in this paper,
it may be possible to extend the class of signals
for Lorentz violation by looking for gravitational analogs 
of the nonminimal electromagnetic sector of the SME,
which goes beyond the minimal $(k_F)^{\ka\la\mu\nu}$ 
coefficients \cite{nonmin}.

\end{document}